# Direct observation of quantized interlayer vortex flow and vortex pinning distribution in high-$T_\text{c}$ La$_{1.87}$Sr$_{0.13}$CuO$_4$ single crystals

I. Iguchi[1], S. Arisawa[1], T. Uchiyama[2], K-S. Yun[1], T. Hatano[1] and I. Tanaka[3]

[1]*Nano System Functionality Center, National Institute for Materials Science (NIMS), 1-2-1 Sengen, Tsukuba, Ibaraki 305-0047, Japan*
[2]*Department of Science Education, Miyagi University of Education, 149 Aramaki-aza-aoba, Sendai, Miyagi 980-0845, Japan*
[3]*Institute of Inorganic Synthesis, Faculty of Engineering, Yamanashi University, 7 Miyamae, Kofu, Yamanashi 400-0021, Japan*

A scanning superconducting quantum interference device (SQUID) microscope (SSM) is used to study the magnetic imaging of dynamic motion of quantized interlayer vortices induced by the Lorentz force in anisotropic high-$T_\text{c}$ La$_{1.87}$Sr$_{0.13}$CuO$_4$ single crystals. It is found that 3 modes of flux motion switch depending on the transport current. By increasing the current a transition from the creep-like behavior of vortices to a steady flow of vortices was observed. Even higher current induced a continuous expansion of vortex-flow area indicating an inhomogeneous distribution of various pinning centers.

As is well known, the quantized vortices in a superconductor experience the Lorentz force and the pinning force in the presence of transport current. When the former is smaller than the latter, the vortices remained to be pinned or creep-like motion occurs. As the Lorentz force exceeds the pinning force, the creep-like motion of vortices changes into a viscous flow characterized by the relation

$$\eta v_L = f_L - f_P \tag{1}$$

where $v_L$ is the vortex velocity, $\eta$ is the viscosity coefficient of medium, $f_P$ is the pinning force and $f_L$ is the Lorentz force given by $f_L = J_T \Phi_0/c$ ($J_T$ : transport current density) [1-3]. The direction of vortex flow is perpendicular to the current direction. A number of measurements on flux flow resistivity [1-3] and application to flux flow devices [4-6] have been reported. In anisotropic high-$T_c$ cuprate superconductors, the flow phenomenon of interlayer vortices induced by a transport current led to novel concept of Josephson plasma terahertz emission [7,8].

Static magnetic images of quantum vortices, fractional vortices [9-19], interlayer vortices [20-22] or vortex-like structure above $T_c$ [23,24] have been reported using a high sensitive SSM technique or a Hall probe technique. While the quantized vortices along the *c*-axis direction appeared as a round shaped image in the *ab*-plane, the interlayer vortices appeared as a long shaped image in the *bc*-plane characterized by a long *c*-axis penetration depth, although they carry one flux quantum $\Phi_0$ in common.

On the other hand, for detection of the dynamic image of flowing vortices, there is no report using SSM or Hall probe techniques. Using a Lorentz microscope technique, Tonomura et al. succeeded in observing the transient behavior of quantized vortices due to a given change of magnetic field in a Nb thin film [25,26]. It was only observable under the condition that the vortices were just released from the pinning sites and the vortex speed was extremely slow.

We here present the first observation of the dynamic flow of interlayer vortices due to the Lorentz force in high-$T_c$ superconductors. The time sequence of flowing vortices with the same polarity in the stationary state yields a time-averaged magnetic signal and it is readily detected by a SQUID provided that the SQUID operation frequency is higher than the vortex flow frequency. The detected SSM flow signal appeared spatially uniformly. The observed results enable us to illuminate the detailed threshold behavior of flux flow phenomenon in a macroscopic scale, for which only speculations based on the current-voltage characteristics have been so far presented.

The high quality $La_{1.87}Sr_{0.13}CuO_4$(LSCO) single crystals were grown by a traveling-solvent floating-zone technique and the typical dimension was about 1mm long, about 3mm wide and 0.2mm thick. The X-ray diffraction pattern showed that the sample crystal was of high quality. $T_c$ of the crystal was 24K. The specimen crystal was polished and set on a substrate so that its *a*-axis direction was normal to the sample surface. With this geometry, the vortices are driven along the *b*-axis direction by the c-axis transport current. The resistance vs. temperature ($R - T$) curve above $T_c$ showed a semiconductive behavior typical of the *c*-axis transport.

The SSM made use of Nb SQUID and had a spatial resolution limited by the pick-up loop of $\phi$10μm. Both the SQUID and the pick-up coil were mounted on a Si chip which was set on a cantilever whose angle against the sample surface was $\theta\sim 5°$. The scanning was done with the tip not touching the sample. The SQUID resolution was $5\mu\Phi_0 Hz^{-1/2}$, and the scanning of sample stage was conducted by three stepping motors along the *x*, *y* and *z* axes. The detector picked up the magnetic flux through the pickup coil nearly normal to the sample surface, i.e. the *z* direction. The SQUID detects the total magnetic flux through the pickup loop. The scanning direction was along the c-axis direction. The use of a permalloy magnetic shield reduced the background magnetic field to be about 0.1μT. A small magnetic field could be generated by the coil wound around the sample. The sample-pickup loop distance was about 3μm.

Figure 1(a) shows a schematic of vortex flow model due to the *c*-axis transport current in an anisotropic high-$T_c$ cuprate superconductor. Figure 1(b) shows an example of the static magnetic image of pinned interlayer vortices (top view) observed in the absence of transport current, together with the quantized vortex image measured independently for a *c*-axis oriented (001) LSCO thin film (Fig. 1(c)). The scanning area was 300x300μm$^2$. In Fig. 1(b), the pinned interlayer vortices with both positive (yellow or orange color) and negative (violet color) polarity are clearly observable along the *b*-axis. Some SSM vortex signals were overlapped due to the spatial resolution of SQUID of ϕ10μm, which does not necessarily mean that the real interlayer vortices overlap. In contrast to the round image of the conventional quantized vortices, the interlayer vortices exhibited an elongated shape image in the *bc*-plane, similar to those of $HgBa_2CuO_{4+\delta}$ [20], $Tl_2Ba_2CuO_{6+\delta}$ [21] and $YBa_2Cu_3O_{7-y}$ [22]. It is found that the isolated vortices carry nearly the integrated flux of (2.2 ±0.3) x10$^{-15}$Wb, which almost corresponds to one flux quantum value $\Phi_0$ (= 2.07x10$^{-15}$Wb). The external magnetic field was about 1μT. This value nearly agreed from the one derived from the number of the pinned vortices in the scanning area, hence *B*= n$\Phi_0$ ≈*H* where n is the density of vortices.

The *c*-axis magnetic-field penetration depth $\lambda_c$ was estimated by using the following London model for the anisotropic superconductor [27,28],

$$B_z(x, y, z = 0, \lambda_c, \lambda_{ab}) = \frac{\Phi_0}{2\pi\lambda_{ab}\lambda_c} K_0(R)$$

$$R = \left(\frac{x}{\lambda_c}\right)^2 + \left(\frac{y}{\lambda_{ab}}\right)^2$$

(1)

where $\lambda_{ab}$ is the *ab*-plane penetration depth and $K_0(x)$ is the modified Bessel function of the $0^{th}$ order. The field is summed over the area of the pickup loop. In the data fitting process, two parameters were assumed. One is the sample-pickup loop distance $z_0$ and the other is the value of $\lambda_{ab}$. From the data of the independent measurement on the isolated quantized vortex, we assumed $z_0 = 3\mu m$ and $\lambda_{ab} \approx 0.1\mu m$. Note that the value of $\lambda_{ab}$ does not affect $\lambda_c$ appreciably since $\lambda_c >> \lambda_{ab}$. To estimate $\lambda_c$, the data fitting process for isolated vortices was done using Eq. (1). $\lambda_c$ was estimated to be $11 \pm 2\mu m$. This value fairly agrees with the one obtained by microwave measurement [29].

Figure 2 shows a series of SSM magnetic images (top view) at different *c*-axis transport currents. For the transport current up to 0.7mA, the distribution of pinned vortices changed only slightly. The slightly different patterns of static images show that the pinned sites were partially changed, which indicates the occurrence of local flux creep motion. For example, the vortex image seen at the position (x = 1180μm, y = -3280μm) in Fig. 2a disappeared in Fig. 2c. A drastic change was observed above 0.8mA. As shown in Fig. 2e, the pinned vortices seen in lower current state disappeared almost suddenly and a flat-plane magnetic image appeared in the right hand side of scanning area. Within this area, the magnetic signal level almost became spatially uniform. We identify this as the occurrence of vortex flow as a result that the Lorentz force overcame the pinning force. The sequence of moving vortices yielded the time-averaged magnetic signal. By changing the transport current 0mA →0.9mA→ 0mA →0.9mA, the sequential change of vortex state: pinned state (P) → flow state (F)

→P →F was confirmed. The magnetic signal height in Fig. 2f was 0.6 - 0.8μT as compared with that of a typical isolated interlayer vortex of about 1.6μT. For the current above 1.0mA, the left front edge of the flux flow region started to move toward left side and covered the whole scanned area for higher currents. The result shows that, with increasing the transport current, the vortices trapped at various pinning sites are released gradually by the increased Lorentz force. The phenomenon was very reproducible and reversible. Note that the flux flow voltage was not detected in this measurement because it was immeasurably small (less than 1nV) due to very dilute density of vortices.

Figure 3(a) shows the cross-sectional views of magnetic images recorded at the $c$-axis transport current $I_T$= 0mA and 1.0mA. The detected flat magnetic signal is clear in this figure. Figure 3(b) shows the cross-sectional view of a wider area of about 600μm at $I_T$= 1mA. It is found that there exist the two flow regions (A and C) and the pinned region (B). Note that the pinned vortices at $I_T$= 1mA were also released for higher currents.

Figure 4(a) shows the plots of magnetic amplitude at a local point in space in the scanning area (x = 1210μm and y= -3530μm) as a function of transport current recorded at the different section of the sample. Initially ( $I_T$ = 0mA), the trapped vortex was absent at this point. With increasing the current, local magnetic amplitude only changed slightly for I< 0.7mA due to flux creep phenomenon. Above 0.75mA, the signal started to rise rapidly, indicating the occurrence of vortex flow. The signal almost saturated between 0.9mA and 1.05mA. In this current interval, as shown in Fig. 4(b), the area of flux flow region increased almost without changing the magnetic signal height, that means, the flow region spread by releasing vortices pinned with different strength from the pinning sites little by little. Figure 4(c) illustrates the observed threshold-current distribution for flux flow as a function of spatial position x. The result shows the pinning force distribution which is not uniform even for the best single crystal.

Above 1mA, the signal dropped rapidly. This happened because the SQUID detection system including the feedback loop could not follow the vortex flow speed $v_L$. $v_L$ is nearly proportional to the current density $J_T$ and depends on the sample crystal medium. It is generally very large for Bi2212 crystals (e.g. $10^6$-$10^7$cm/sec for $J_T = 10^4$-$10^5$A/cm$^2$), whereas it is small for LSCO crystals (e.g. 5x10$^3$cm/sec for $J_T = 3.6$ x10$^2$A/cm$^2$ [30]). The effective transport current density in our measurement was $J_T \sim$ 1A/cm$^2$ at $I_T$= 1mA, hence we estimate $v_L \sim$ 15cm/sec by assuming the same viscosity coefficient of LSCO medium. The practical SQUID detectable frequency is about 3kHz, which yields the detectable vortex velocity to be about 10cm/sec by considering the observed vortex number density of 5x10$^4$/cm$^2$. The order of estimate agrees with the experimental observation.

The magnetic signal height in the vortex-flow state is also estimated as follows. We simulate the moving vortices as the pulse trains (period: $T$) of the form $\Phi(t) = \Phi_p \cos^2(\frac{\pi}{\tau})t$ where $\Phi_p$ is the value of magnetic flux at the peak point and $\tau \approx 2\lambda_c/(1-e^{-1})v_L$, then the time averaged magnetic flux is given by

$$\langle \Phi(t) \rangle = \frac{1}{T}\int_t^{t+T} \Phi(t)dt = \frac{1}{T}\int_t^{t+T} \sum_{n=0}^{\infty} \frac{\Phi_p \tau}{n\pi} \frac{\sin\frac{n\pi\tau}{T}}{[1-(\frac{n\tau}{T})^2]} \cos\frac{2\pi nt}{T} dt = \frac{\tau}{2T}\Phi_p \qquad (2)$$

The period $T$ is given by $T = (n_0 v_L)^{-1}$ where $n_0$ is the number density of vortex per unit length, hence $\langle \Phi(t) \rangle \approx \frac{e}{e-1} n_0 \lambda_c \Phi_p$. For the vortices with $\Phi_p$=1.6μT, $\lambda_c \approx$11μm and its number density of 5x10$^4$/cm$^2$, the time-averaged magnetic signal height is estimated to be about 0.6µT, in reasonable agreement with the observed value of 0.6 - 0.8µT.

In case that the transport current is perpendicular to the *c*-axis, such vortex flow phenomenon was not observable. Only a slight change of pinned vortex distribution was observable. The possible contribution of transport current to the z-component of magnetic signal near the center of scanning area was estimated to be at least less than 0.01µT, which is negligibly small.

Consequently, we have succeeded in visualizing the flux motion of the vortices in a LaSrCuO single crystal by scanning SQUID microscopy. The real spatial images revealed a non-uniform distribution of various pinning centers. Depending on the c-axis transport current, a transition from the creep-like behavior of vortices to continuous band-like magnetic images above a certain threshold current was observed, indicating the occurrence of a steady flow of vortices. For further increase of current, the continuous expansion of vortex-flow area was also recognized, demonstrating the distribution of the pinning forces. The sequential release effect of vortices from the pinning sites showed the threshold current distribution of about a few tens of percent due to the pinning centers with various pinning strength. Moreover, the observation of elongated vortices and large anisotropy in the penetration depth suggests more 2D-likebehavior than 3D-like [31].

The authors are very grateful to Dr. X. Hu for invaluable discussions.

References

[1] Y. B. Kim, C. F. Hempstead, and A. R. Strand, Rev. Mod. Phys. **36**, 43(1964).

[2] C. F. Hempstead and Y. B. Kim, Phys. Rev. Lett. **12**, 145 (1964).

[3] Y. B. Kim and M. J. Stephen, in "Superconductivity" Volume 2 edited by R. D. Parks (Marcel Dekker Inc. 1969), p.1107.

[4] T. Nagatsuma, K. Enpuku, F. Irie and K. Yoshida, J. Appl. Phys. **54**, 3302(1983).

[5] S. N. Erne and R. D. Parmentier, J. Appl. Phys. **52**, 1091 (1981).

[6] J. U. Lee, J. E. Nordman and G. Hohenwarter, Appl. Phys. Lett. **67**, 1471 (1995).

[7] E. Kume, I. Iguchi and H. Takahashi, Appl. Phys. Lett. **75**, 2809(1999).

[8] M. Tachiki, M. Iizuka, K. Minami, S. Tejima and H. Nakamura, Phys. Rev. B **71**, 134515 (2005).

[9] C. C. Tsuei *et al.*, Phys. Rev. Lett. **73**, 593 (1994).

[10] J. R. Kirtley *et al.*, Phys. Rev. Lett. **76**, 1336 (1996).

[11] C. C. Tsuei and J. R. Kirtley, Rev. Mod. Phys. **72**, 969 (2000).

[12] T. Morooka, S. Nakayama, A. Odawara and K. Chinone, Jpn. J. Appl. Phys. **38**, L119 (1999).

[13] A. Sugimoto, T. Yamaguchi and I. Iguchi, Physica C **367**, 28 (2002).

[14] H. J. H. Smilde *et al.* Phys. Rev. Lett. **88**, 057004 (2002).

[15] H. Hilgenkamp *et al.*, Nature **422**, 50 (2003).

[16] A. M. Chang et al., Appl. Phys. Lett. **61**, 1974 (1992).

[17] A. Oral, S. J. Bending and M. Henini, Appl. Phys. Lett. **69**, 1324 (1996).

[18] A. Oral, S. J. Bending and M. Henini, J. Vac. Sci. Technol. B **14**, 1201 (1996).

[19] I. Maggio-Aprile, Ch. Renner, A. Erb, E. Walker and Ø. Fischer, Phys. Rev. Lett.

**75**, 2754 (1995).

[20] J. R. Kirtley, K. A. Moler, G. Villard and A. Maignan, Phys. Rev. Lett. **81**, 2140 (1998).

[21] K. A. Moler, J. R. Kirtley, D. G. Hinks, T. W. Li and M. Xu, Science **279**, 1193 (1998).

[22] I. Iguchi, T. Takeda, T. Uchiyama, A. Sugimoto and T. Hatano, Phys. Rev. B **73**, 224519 (2006).

[23] I. Iguchi, T. Yamaguchi and A. Sugimoto, Nature **412**, 420 (2001).

[24] A. Sugimoto, I. Iguchi, T. Miyake and H. Sato, Jpn. J. Appl. Phys. **41**, L497 (2002).

[25] T. Matsuda, K. Harada, H. Kasai, O. Kamimura and A. Tonomura, Science **271**, 1393 (1996).

[26] N. Osakabe, H. Kasai, T. Kodama and A. Tonomura, Phys. Rev. Lett. **78**, 1711 (1997).

[27] J. R. Clem and M. W. Coffey, Phys. Rev. B **42**, 6209 (1990).

[28] J. R. Kirtley, V. G. Kogan, J. R. Clem and K. A. Moler, Phys. Rev. B **59**, 4343 (1999).

[29] T. Shibauchi *et al.*, Phys. Rev. Lett. **72**, 2263 (1994).

[30] T. Tachiki, Ph.D. thesis, Tohoku University (2000).

[31] U. Divakar *et al.*, Phys. Rev. Lett. **92**, 237004 (2004).

Figure captions

FIG. 1 (a) Schematic of vortex flow phenomenon induced by transport current in an anisotropic high-$T_c$ superconductor. (b) SSM magnetic image of the surface of a (100)$La_{1.87}Sr_{0.13}CuO_4$ single crystal without transport current. (c) SSM magnetic image of a *c*-axis oriented LaSrCuO thin film. In contrast to the round image of quantized vortices in (c), the interlayer vortices appeared as an elongated vortex image in (b).

FIG. 2 A series of magnetic images (top view) of the surface of (100) oriented $La_{1.87}Sr_{0.13}CuO_4$ single crystal at different transport currents at 3.2K in the scanning area of 300x300μm$^2$. The transport currents for **a – h** are 0mA, 0.2mA, 0.5mA, 0.7mA, 0.9mA, 1.0mA, 1.06mA and 1.15mA, respectively. With increasing the transport current, the observed magnetic images only slightly changed below 0.7mA due to flux-creep phenomenon. At 0.8mA, a continuous band-like magnetic image suddenly appeared, indicating the occurrence of flux flow motion.

FIG. 3 (a) SSM magnetic images (cross-sectional view) at transport currents $I_B$= 0mA and 1.0mA for the LSCO single crystal. The flat magnetic signal which indicates the occurrence of flux flow is evident for $I_B$= 1.0mA image. (b) The SSM magnetic image (cross-sectional view) in the wider region (600μm) at $I_B$= 0.9mA. The A and C regions correspond to the flux flow region, while the B region corresponds to the flux pinned region. For higher currents, the B region also converted from the pinned state to the flow state.

FIG. 4 (a) Local magnetic amplitude at a specified point (x = 1210μm and y= -3530μm) in the scanning area as a function of c-axis transport current. (b) SSM images (top view) **a – h** correspond to $I_T$ = 0.80, 0.86, 0.92, 0.96, 0.98, 1.02, 1.04, 1.15mA, respectively. (c) Threshold current for flux flow as a function of the spatial position *x* in the sample crystal. The horizontal bar represents the width of scattered data at the threshold in the region of 130μm< x <300μm. The result indicates that the pinning force is distributed inhomogeneously in the sample and the depinning process occurs little by little with increasing transport current.

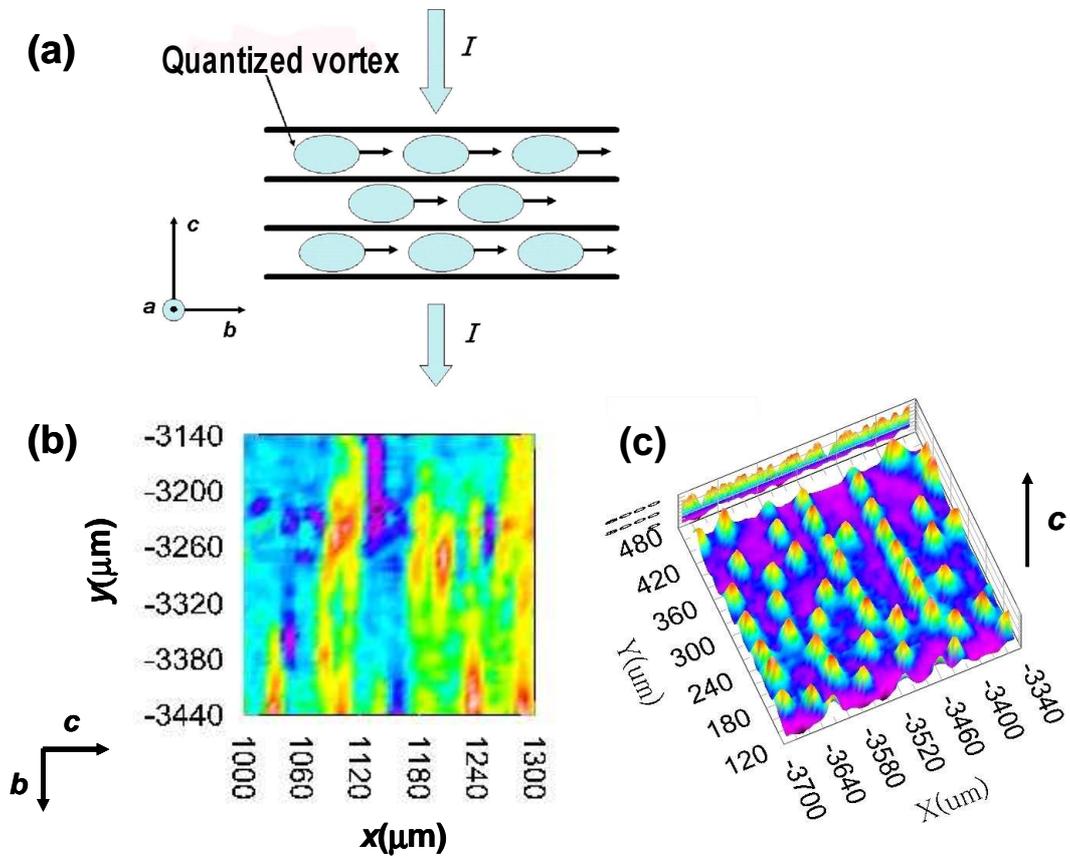

FIG. 1

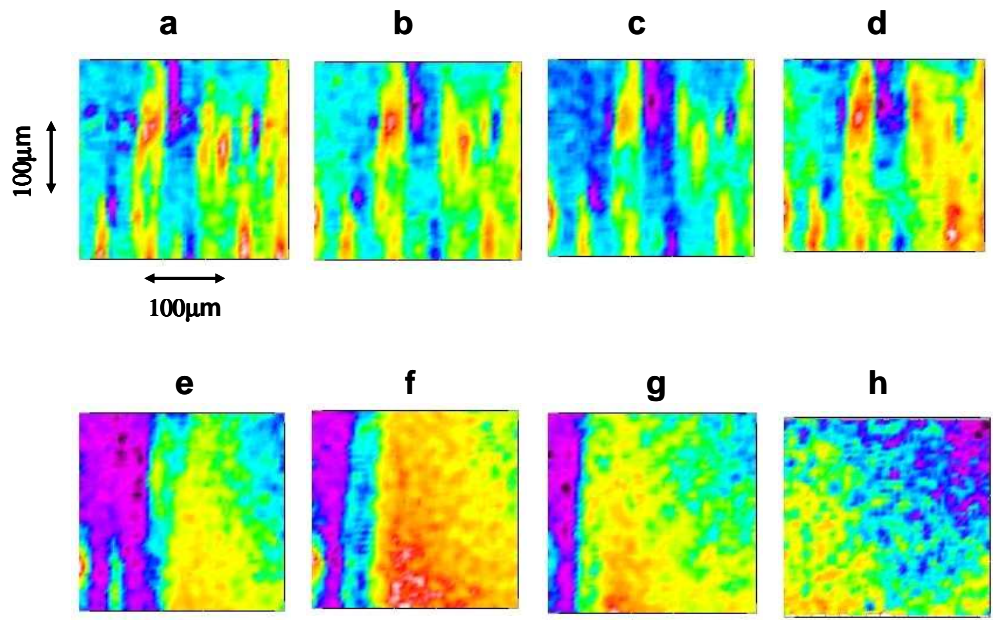

FIG. 2

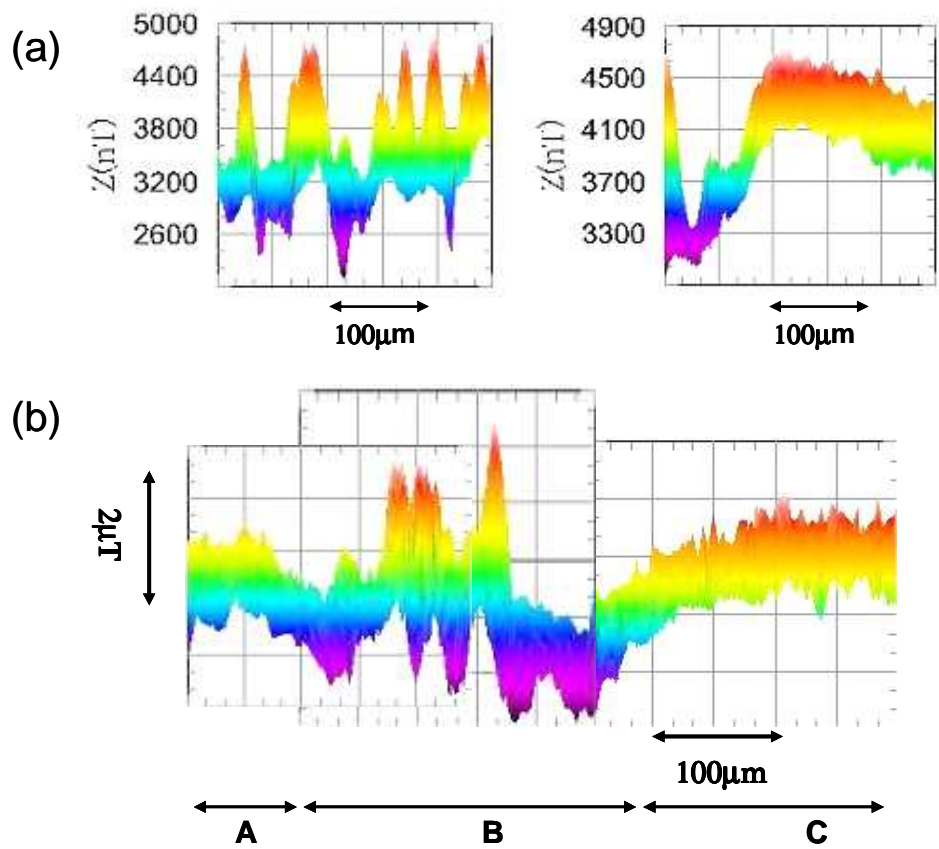

FIG. 3

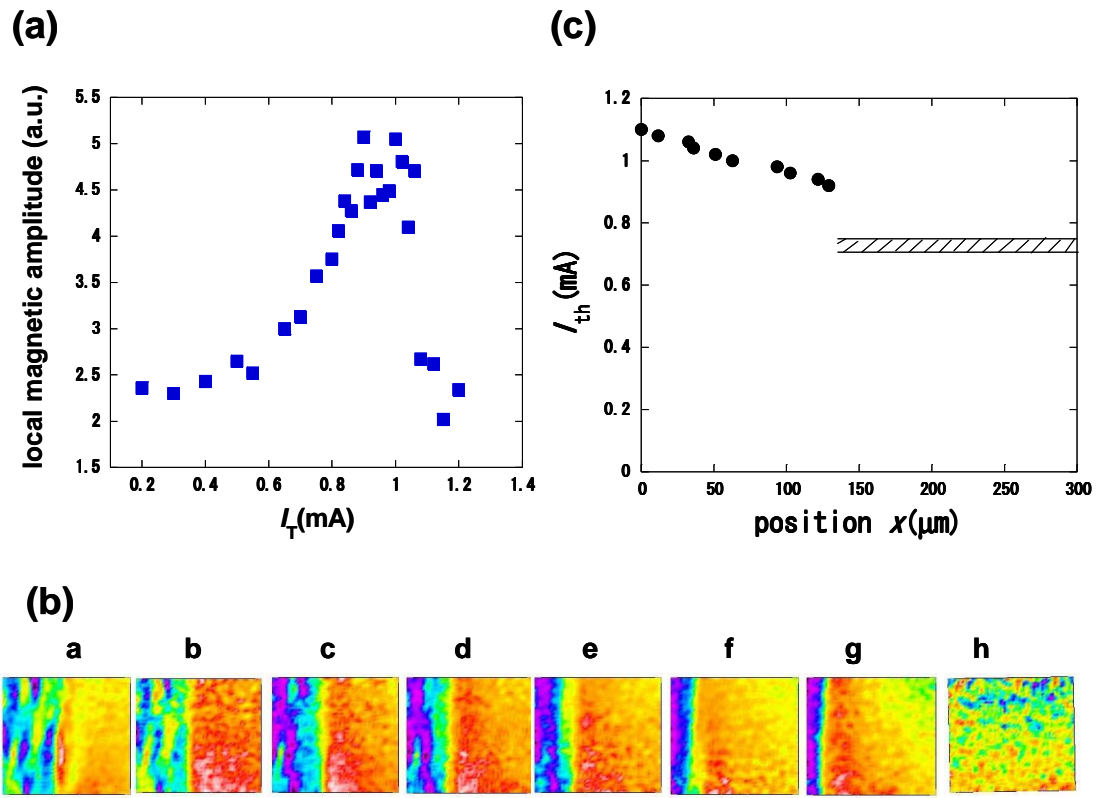

FIG. 4